# Adopting a software product line engineering approach in industrial development contexts: A protocol for a systematic literature review

José L. Barros-Justo[1], Luisa Rincón[2,4], Ángela Villota[3,4], Wesley K. G. Assunção[5]

[1]*School of Informatics (ESEI), Universidade de Vigo, Ourense 32004, Spain. (jbarros@uvigo.es)*
[2]*Pontificia Universidad Javeriana, Cali, Colombia. (lfrincon@applies.variamos.com)*
[3]*Universidad Icesi, Cali, Colombia. (apvillota@icesi.edu.co)*
[4]*Centre de Recherche en Informatique (CRI), Paris 1 University, Paris, France,*
[5]*Federal University of Technology - Paraná, Toledo, Brazil. (wesleyk@utfpr.edu.br)*

**Abstract:**

The value of a systematic secondary study (a systematic mapping study (SMS) or a systematic literature review (SLR)) comes, directly, from its *systematic* nature. The formal, well-defined, objective and unbiased process guarantees that the results from these systematically conducted studies are valid. This process is embodied in an action protocol, which must be agreed upon by all the researchers before conducting the secondary study. The protocol is, therefore, a detailed action plan, which contains all the tasks and the ordered sequence of steps to be executed. This document details that protocol for a SLR on the adoption of the software product line engineering (SPLE) approach in industrial development contexts. The goal of that SLR is to identify and analyse the benefits and drawbacks that this adoption has had in industrial development contexts, in contrast to the experiences reported in academic environments.

## 1. Introduction

*"EBSE (evidence-based software engineering) is concerned with determining what really works, when and where, in terms of software engineering practice, tools and standards"* [1]. The two key tools of EBSE are systematic mapping studies (SMS), also known as scope studies and systematic literature reviews (SLR). The main goal of an SMS is to build a classification schema for the topics of interest and to provide an overview of a research area (the map), while a SLR focus on aggregating all empirical studies on a particular topic to synthesize new knowledge [1,2].

One of the most frequent applications of a SLR is the collection of evidence to identify, analyse and interpret the reported evidence. The objective is to respond to a set of research questions by analysing the data extracted from primary or secondary sources, and the synthesis of new information, in an unbiased way, through a repeatable (to a certain extent) process. The identification of gaps (lack of studies) or inconsistencies among results are other frequent goal of SLRs.

Since its inception, around 2004, the methodology associated with the conducting of systematic secondary studies has evolved, refining individual processes, and offering guidelines that included the best practices. The last and most cited specific guide for SMS is reported in [2] and for SLR in [1,3,4].

Although the literature offers a significant number of secondary studies in the area of software product line engineering (SPLE), information on the benefits that this practice have provided to the industry and the problems, inconveniences or costs caused by its adoption, is very scarce. A preliminary investigation allowed us to discover that a large part of the available literature on SPLE referred to proposals for solutions to specific problems, identified in academic environments, with a weak practical component. Since our goal is "*to determine what really works, in terms of software engineering practice*", we decided to conduct a SLR to analyse the adoption of SPLE approaches by industries and the impact it caused (benefits and drawbacks). For this purpose, we present a detailed protocol in the following sections.

## 2. Protocol

A protocol is a document that describes, in detail, the way in which a research activity will be conducted. In particular, the protocol describes the activities to be carried out, their temporal sequence, those responsible for each task, the necessary resources, the control mechanisms and the range of acceptable output values.

Being a document prior to the execution of the study, the protocol is the main cause of the systematic nature of the review. The benefits of a good protocol include [1]:

- allow the replication of the original study, by offering all the necessary information to repeat the processes, including structural and temporal aspects (context),
- reduce the bias of researchers, by formalizing and controlling the activities in which they are involved (reduction of threats to validity),
- estimate the effort necessary to carry out the study,
- As a side effect: reduce the length (number of pages) of the systematic study to be published, offering the protocol as an independent document that can be cited from within the secondary study.

Like any plan, a protocol must be flexible and adaptable to situations that may arise during the execution phase of the study. If the deviations from the study with respect to the original protocol are significant, they have to be reported in the final study (the published work).

The following subsections (2.1 to 2.7) detail the key activities covered by the protocol for the subsequent execution of the SLR.

### 2.1. Goal and Research Questions

The first step in a systematic study is to establish the general objective. The goal of the study will help to determine how we are going to summarize and synthesize inputs from different studies. For example, our study is aimed to assess the adoption of a technique in industry, therefore, it is likely that we will gather and synthesize the results of previous observational studies, case studies and experience papers, classifying the outcomes as benefits or drawbacks.

This study will address questions about the application of a specific technology (SPLE), we are not interested in comparisons with other technologies, but to gather knowledge related to the effects of the adoption of SPLE, that means that our SLR will be a qualitative review. Our interest is in studying adoption issues (associated costs, obtained benefits and so on).

Under the above considerations, the Goal of this study can be established as:

*To identify and analyse the benefits and drawbacks that adopting a software product line engineering approach has had in industrial development contexts.*

To this purpose, we establish the following set of research questions (RQs):

- *RQ1: What approaches to the adoption of SPLE in industrial environments have been reported?*

    **Rationale:** literature reports on three different approaches to develop software product lines [5]:

    1. The proactive approach: from scratch, by applying a complete domain analysis and variability management before any application engineering,
    2. The reactive approach: by creating and updating the SPL when every new product appears, and
    3. The extractive approach: which takes existing products to extract common and variable assets.

    In this RQ we want to investigate which approach is more common (frequency) in industry and the potential relationships between the approach and the drawbacks/benefits identified.

- *RQ2: Which drawbacks have been reported related to the adoption of SPLE?*

    **Rationale:** To produce a map (or taxonomy, or a list) of the most frequently reported drawbacks (such as issues, costs, risks, downsides or in general any negative outcome) experienced by SPLE adopters. Some of these drawbacks are, for example: organizational cost, cost of developing the core assets, cost of update the core assets, cost of reuse, and many more [6].

- *RQ3: Which benefits have been reported related to the adoption of SPLE?*

    **Rationale:** To identify the most frequently reported benefits, for example: cost-savings, shorter time-to-market, user engagement or satisfaction, a larger catalogue of products, maintenance effort reduction or project quality improvement [7].

- *RQ4: What are the main characteristics of the developed SPL?*

    **Rationale:** This RQ is aimed to "contextualize" the experience. We are interested in factors such as quantity of derived products, development team size, how long has the SPL been "active", the industrial context in which the experience took place, etc.

- *RQ5: Which techniques, methods and metrics have been used to assess the drawbacks and benefits related to the adoption of SPLE?*

    **Rationale**: To identify which method or technique have been used, e.g.: COPLIMO, INCOME, SIMPLE or SOCOEMO-PLE, and the metrics more frequently considered, e.g.: money, time, people, hardware, infrastructure, training, number of requirements, ROI, NPV, etc.

This study will be framed by a time period from the inception of the SPL in 1980 [8] to March 2019, (we will run the first search on first of April 2019).

We also complement this systematic review by including some Demographic Questions (DQs) frequently reported in secondary studies, such as:

- DQ1: What are the most relevant publishing venues?
- DQ2: Who are the most active researchers, organizations and countries?
- DQ3: What has been the evolution of the number of publications over the years?

DQ1 is useful for researchers interested in publishing related work. DQ2 helps identify potential biases, for example, in the situation in which a particular author or organization is over-represented in the set of selected papers. Finally, DQ3 allows observing the evolution of interest in the research area and inferring its future trend. A mature area will have a stable evolution, or descendant, in the number of publications. While an emerging area, with high interest (a hot topic), will show a growing number of publications over time.

## 2.2. Search strategy

Following the guidelines proposed in [1,9–15], the search strategy combines three complementary searches: manual search, automated search, and snowballing.

### 2.2.1 Manual search

The purpose of this manual search is to provide a base set of publications useful for refining and validating the search strings for the automated search strategy, later on (See Subsection 2.2.2).

The manual search will focus on:

- The proceedings of the two most relevant conferences in the area of interest: Software Product Line Conference (SPLC), and International Conference on Software Reuse (ICSR),
- The Empirical Software Engineering Journal (ESE), a venue that frequently publishes industrial experience reports and studies with a strong empirical component,
- Two books from renown authors in the area of SPL [5,16] and
- A web page that gathers the most representative organizations in the area, with outstanding contributions (SPL projects) [17].

The manual search process can be divided into the following steps:

1. Retrieve all the publication from the identified sources on the first of April, 2019,
2. Fill an spreadsheet (Manual_Search.gsheet) with all the raw data (Metadata including: Title, Authors, Venue, Publishing date, Abstract and so on) from step 1,
3. Two sets from the identified papers will be produced (H1 & H2), with an overlap of 50% (i.e. H1 and H2 share a 50% of identical papers),
4. A pair of authors (P1), collaboratively, will select the papers from H1 (applying the selection criteria in Section 2.3),
5. Other pair of authors (P2), will do the selection from H2,
6. First author will integrate the results from the above two pairs of authors,
7. Meetings will be held to deal with possible disagreements. The criteria to decide if a paper should be included or not is explained in Section 2.3.,

8. The set of selected papers will pass to the quality assessment (QA) process and, the papers that pass it, will go to the extraction of data.

The estimated time to perform this search strategy is 2 weeks.

### 2.2.2 Automated search

The automated search will include three activities:

1. Select the Electronic Data Sources (EDS) to be used,
2. Build the search string and adapt it to the query format of every EDS,
3. Run the searches in the selected EDS

To avoid bias we will include four complementary EDS which are well known among researchers and academics: ACM Digital Library (ACM DL), IEEE Xplore, SCOPUS and Web of Science, as suggested by Bailey et al. in [10]. The first two EDS covered the most important journals and conferences in the field of software engineering [18], while the last two are recognized as the largest general indexing services, including papers published by ACM, IEEE, Elsevier, Springer and Wiley.

The building of the search string will be a refining, iterative process. We will include the keywords used in the set of previously selected papers from the manual search strategy. Furthermore, key terms from the Goal of our study, and the RQs and DQs, will be included as well. Finally, synonyms of the previous terms will be reviewed and tested for inclusion. The PICO technique will be used to build up the final search string [19].

Pilot searches will be conducted in SCOPUS to refine the search string until we achieve the following two goals:

1. The set of retrieved papers is manageable (less than 3,000 results),
2. The automated search is able to retrieve, at least, 95% of the papers retrieved by the manual search.

Finally, we will tailor the search string to every EDS. The adapted search strings and the results of the searches will be reported in a table similar to Table 1.

*Table 1        Tailored search strings and number of retrieved documents*

| Search string | EDS | Results |
|---|---|---|
| | ACM DL | |
| | IEEE Xplore | |
| | Springer | |
| | SCOPUS | |
| | WoS | |
| | **Total =** | |

As in the manual search strategy, the results (raw data) from the automated search will be stored in a spreadsheet (Automated_Search.gsheet) with the following organization:

1. First Tab will contains general information about the search: Date and a copy of Table 1,
2. Second Tab will store the raw data from the first EDS (ACM DL),
3. Third Tab will store the raw data from IEEE Xplore,
4. Fourth Tab will store the raw data from SCOPUS,
5. Fifth Tab will store the raw data from WoS,
6. Sixth Tab will integrate raw data from all the EDS and Highlight the Duplicates entries

The estimated date to run the automated search is April, 15th.

### 2.2.3 Snowballing

The first author will perform a forward and backward snowballing search process as described in [15]. To prevent the uncontrolled growing of identified studies during the snowballing process, a refined set of initial seeds will be chosen, and special care will be taken to select the seeds for every iteration of the process. The initial seeds will be the papers that go through the selection, quality assessment and data extraction processes. This implies that the snowballing search will be the last search process to conduct (using the final set of selected papers as seeds for the first iteration).

All the data for every iteration of the process will be stored in a spreadsheet (Snowballing_Search.gsheet), including seeds, references and citations, and the criteria used to exclude the papers that will not go through to the next iteration (selection criteria, Section 2.3). Every iteration of the process will be stored in a separate Tab of the spreadsheet.

The estimated date to run the snowballing search is July 2019.

### 2.3. Selection process

In order to select the articles, we will follow the selection process depicted in Figure 1 and suggested in [20,21]. The input to the selection process is the "data set universe", which in turn is composed of the metadata of references found by the two first search strategies (manual and automated). The snowballing process will store its own data related with the selection process (spreadsheet Snowballing_Search.gsheet). During the process, three temporary data sets will be built:

- *excluded references*: stores references excluded by any activity during the selection or the data extraction process;
- *borderline references*: has those references for which a revision of the full text should be done, because a decision to include or exclude the reference has not yet been reached; and
- *candidate references:* keeps references that can be moved to the set of selected papers.

These three sets will be stored in one spreadsheet (Selection.gsheet) with three separate Tabs. The output of the selection process will be the input for the Quality Assessment process (Section 2.4).

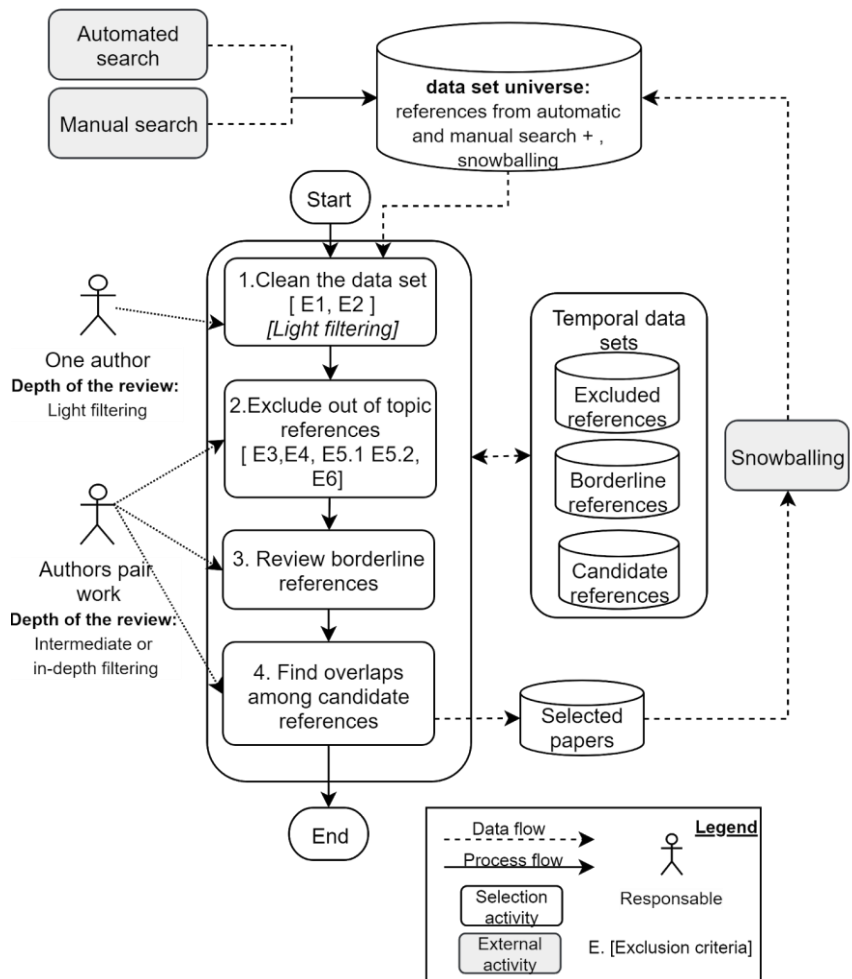

*Figure 1*                    *Selection process*

Four activities conform the selection process: *(1) clean the data set*; *(2) exclude out of topic references; (3) review borderline references; a*nd (4) *find overlaps among candidate references.*

In the first two activities the references will be filtered using six exclusion criteria, hereafter identified from E1 to E6 (see Table 2). Initially, references will be filtered by title and venue. If this information is not sufficient, keywords and abstracts will be reviewed and if this information remains insufficient, the full text of the article will be reviewed. For example, exclusion criteria E1 and E2 only require a light revision of the metadata, while exclusion criteria E3 to E6 require a more in-depth revision, as they are more related to the content of the paper.

The filtering of the studies follows a cascade-style process, papers not excluded by criteria E1 serves as inputs for E2, while the output from E2 serves as input for E3 and so on. However, it is still possible to exclude papers in any activity of the selection process, or even in the data extraction process. This may occur when a paper fails to meet some of the exclusion criteria in a later stage. For example, suppose that a paper whose metadata is in English passes all the filters, but once we download it we find that its content is in another language. This paper does not satisfy E2 anymore and therefore should be excluded.

In order to preserve the full traceability of the selection process, a record will be kept to document the exclusion criteria that explains why a paper was excluded. This record will be stored in the spreadsheet Selection.gsheet, where six columns named E1 to E6 will be added to the raw data (data set universe). A mark (an X) under one of these columns means that the paper in that row satisfies the exclusion criterion and should be discarded.

| Exclusion criteria | Procedure |
|---|---|
| E1. Duplicated papers. i.e., references found more than once during the search processes.<br>  E2. Not written in English | Light filtering. Based on title and venue and number of pages.<br>Source: reference meta-data |
| E3. Posters, books, dissertations, tutorials, slides, panels and any piece of work that can be considered as grey literature.<br>E4. Secondary studies (they, if any, will be considered in our Related works section).<br>E5. The focus of the paper is not on the application of SPLE in real world environments.<br>  E5.1.The focus is not on SPLE.<br>  E5.2.The focus is not on SPLE adoption in real world environments (pure academic experiences will be excluded).<br>E6. The paper do not report on any benefit or a drawback due to the adoption of the SPLE approach. | Intermediate filter.<br>Based on title. If available, keywords and abstract will be checked too.<br>Source: reference meta-data<br><br>[Optional] In-depth filtering. Sections such as introduction, methods and conclusions will be checked. If need the full paper content will be reviewed.<br>Source: downloaded paper |

In the third activity of the selection process, references in the "*borderline references*" dataset will be checked until there are no references left. To do this, the articles will be downloaded and their contents will be reviewed by teams. Finally, in the fourth activity, we will review the references from the "*candidate references*" set that have similar titles or authors. This review aims to filter overlaps between publications to avoid double counting of study results. Two cases are possible here: that a paper reports multiple studies or that a study is reported in more than one paper [1]. In the first case, each study reported in the paper will be considered as a separate study for the purposes of this systematic review. In the second case, the most complete paper will be selected. Another scenario occurs when the same research is reported in different articles and each article shows a different part of the research. In this case, all articles will be selected.

Regarding those responsible for carrying out the activities, the first activity will be carried out by only one of the authors since, as shown in Table 2, the exclusion criteria for this activity requires just a light revision (E1 to E2). For the rest of the activities, the authors will work in pairs to reduce bias. Each pair of authors will place the references in the corresponding dataset according to the following rules ( inspired by those in [2]):

- *Candidate references*: both reviewers agree on the inclusion of the reference.
- *Excluded references*: both reviewers agree on the exclusion of the reference or one reviewer excludes it whereas the other has doubts.
- *Borderline references*: both reviewers have doubts or one reviewer includes the reference whereas the other has doubts. Doubts in the review of references may occur for example when the meta-data do not provide enough information to make the decision or because one of the reviewers wants another opinion.

### 2.4. Quality assessment

The purpose of including a quality assessment (QA) process can serve the following objectives:

1. Determine if the differences in the results can be explained due to differences in the quality of the included primary studies,

2. Ponder the importance of the different individual contributions in relation to the overall results,

3. Reduce the effort in the data extraction phase, by including only those primary studies that exceed a pre-established threshold (minimum level of quality).

As in the selection process, two important decisions must be made during the QA:

1. The criteria to be used to evaluate quality (What?), and

2. The evaluation process (How? & Who?)

For the criteria we have adapted the checklists suggested by Chen & Babar [22] and Steinmacher et al. [23] to include the industrial aspect (QA2). Unlike other studies in which quality is evaluated according to the type of study or the applied research method, in our case, since these are studies that report experiences in the industry, we will use other quality indicators, closer to the practice. Each of the criteria can result in one of three possible responses: Y(es), P(artial) and N(o), with a quantitative assessment of 1, 0.5 and 0, respectively (Table 3). Therefore, the maximum score a paper can reach is 8 (all QAs valued at 1) and the minimum is 0. Only those papers that reach a rating of 5 will go through the data extraction process.

*Table 3       Quality assessment criteria*

| | **Criteria** |
|---|---|
| QA1 | Is the purpose of the study clearly defined? (The goal, objective, aim…) |
| QA2 | Is the context adequately described? (type of industry, domain, size, number of participants, department or section, country or region ...) |
| QA3 | Is the method of data collection reported? (interviews, surveys, direct observation, direct participation, software tools ...) |
| QA4 | Is a sufficient amount of data reported? (to answer our RQs) |
| QA5 | How is the analysis and interpretation of the data reported? (method, who, when, tool, completeness and rigor) |
| QA6 | Are the results clearly reported? |
| QA7 | Is there a clear path from the evidence (data) to the conclusions? |
| QA8 | Are threats to validity explicitly reported? (for example: bias) |

The output of the Selection process will be the input for the QA process. A couple of reviewers will carry out the QA process independently. The first author will integrate the results of each reviewer. Possible discrepancies will be treated in meetings with both reviewers, until reaching an agreement.

This QA process will be performed in parallel with Data extraction, but giving priority to QA, i.e. only those papers that go through the QA would be considered for Data extraction.

### 2.5. Data extraction

We created a Data Extraction Form (DEF) to objectivize the data extraction process (Figure 2). The use of a DEF helps us to keep the consistency and facilitates the integration of the results. It is possible that the DEF undergoes minor modifications during the conducting phase of the SLR, due to the type and amount of data found. Another section of the DEF, which is not shown in Figure 2, contains de results for the DQs, other frequent data in secondary studies, such as venue, year, authors and affiliations, etc.

The DEF was implemented in a spreadsheet format. Columns store the data needed to answer every RQ of our study, while rows represent the reviewed studies. Several contiguous columns can belong to the same RQ. A reviewed article may occupy more than one row. Every cell contains a text (Data) that provides enough data to answer a specific item for a research question (ItemX_RQj). A comment will be attached to every cell, with text extracted from the original paper, to support the data. Every group of columns associated to a specific RQ has a different colour, to ease the visualization of related data.

*Figure 2*          *Data Extraction Form (DEF)*

To reduce potential bias, the extraction process will be conducted as follows:
1. The set of selected papers will be split into two halves: H1 and H2,
2. A pair of reviewers (say R1 and R2) will extract data from H1, independently,
3. Another pair of reviewers (say R3 and R4) will extract data from H2, independently,
4. Meetings will be held to resolve any arising discrepancy.

Table 4 below shows the RQs and DQs, and the piece of data that should be extracted to answer them.

| Question | Piece of data | Description |
|---|---|---|
| RQ1 | SPLE approach | Valid inputs are proactive, reactive, or extractive. |
| RQ2 | Drawbacks | Collect data about any negative outcome. For example: organizational cost, cost of developing the core assets, cost of update the core assets, cost of reuse, additional development cost, cost of producing unique artefacts, low quality of the (derived) sub products, developer's stress, effort increase, more complexity, lack of supporting tools... |
| RQ3 | Benefits | Collect data about cost-savings, shorter time-to-market, user engagement, satisfaction, a larger catalogue of products, maintenance effort reduction, project quality improvement. |
| RQ4 | Characteristics of the developed SPL | Collect data about quantity of derived products, team size, how long the SPL has been "active", market segment or domain, and any other "features" help to describe the SPL. |
| RQ5 | Methods or techniques | For example: COPLIMO , INCOME, SIMPLE, SOCOEMO-PLE, COCOMO, ad-hoc or no method. |
| | Metrics | Collect data the metrics applied to measure the effect. For example: money, time, people, hardware, infrastructure, training, communication, LoC, number of requirements, ROI, NPV... |
| DQs | PaperID | A code to uniquely identify the paper along the processes of our study. |
| | Title | Collect the publication title. |
| | Venue name | Collect the name of the publication venue. |
| | Venue type | Three valid types: Journal, Workshop, Conference. |
| | Authors Info | Collect author names, Organizations, Countries. |
| | Publication Year | Collect the year of the publication. |

## 2.6.   Validity threats

The validity threats can be classified as descriptive validity, theoretical validity, interpretive validity and repeatability [24,25].

**Descriptive validity**. This threat is about the extent to which observations are described accurately and objectively [2]. This threat will be reduced by the use of a data extraction form (DEF) designed to register and maintain traceability of the data extracted from the selected papers.

**Theoretical validity.** This threat refers to the capacity of the research of capturing what we intend to capture [2]. A secondary study is by definition limited by the search date, the electronic data sources and the key terms used in the search. We will use the following strategies to reduce this threat.

First, we had several discussions to refine the goal and research questions of our search, so we argue the final set of research questions truly reflect the goal of our work.

Second, one of the major concerns in systematic reviews is finding all the relevant primary studies (evidence). In this case, we will use three complementary search strategies to ensure we find the largest number of related studies: manual search, automated search and snowballing.

- Manual search: we will review, the proceedings of the two major international conferences related to SPLE and software reuse (SPLC and ICSR) since their earliest editions. From these sources we will select those papers that relate to the goal of our study and provide data to answer the research questions.
- Automated search: we will run queries in four electronic data sources (EDS) covering the major publishing venues in the area of software engineering, as suggested in [1,12,26]. To find the maximum number of primary papers, the quality of the search string will be refined iteratively until it reaches a good level as was explained in Section 2.2.2. However, given that some studies that used different terminology to describe their content may be lost, the other two search strategies (manual and snowballing) will help to overcome this drawback, as they are not based on search strings.
- Snowballing: we will use backward and forward snowballing, using the previously selected papers as seeds.

Finally, regarding the quality of data, we plan to exclude grey literature. We assume that good quality papers would appear as journal or conference papers only.

**Interpretive validity:** it is achieved when the conclusions drawn are reasonable given the data [8]. A threat in interpreting the data is researcher bias. To reduce this threat the authors agreed a preliminary review protocol, defined a data extraction form and a process for extracting consistent and relevant information, and will check whether the data to be extracted would address the research questions. Moreover, as the crosscheck is necessary among the reviewers, we will have at least three researchers extracting data independently. Authors will work in pairs to reduce bias. Each pair will independently resolve any divergence or disagreement that arises during the selection process, as it is explained in Section 2.3.

**Repeatability.** There is always a risk in replicating a study and find similar and consistent results [2]. We will mitigate this threat by providing this document, which is the detailed description of the research protocol. Moreover, any additional source of information such as spreadsheets with the results of the searches and the selection process, the DEF, the bibliographic references, etc., will be make publicly available online.

### 2.7. Reporting

Throughout our work (the SLR), we will try to follow the recommendations on good information practices offered by Budgen et al. in [3]. We will also use the reporting structure suggested by Kitchenham et al. in [1]: *Abstract, Introduction, Background, Method, Results, Discussion (including Validity Threats) and Conclusions.*

The space restrictions in a printed publication prevent the presentation of all the necessary data to ensure the total replicability of the systematic study. For this reason, we will provide an online resource with all the key data, to allow for a replication of the study, including the protocol (this document), pilot searches, spreadsheets with selection results, the data extraction form and the extracted data, plus tables and related graphs.

Some good practices to report results propose the following ordered sequence (structure) for every RQ:

1. summary of the most relevant results,
2. a graph or a table (or both, if needed) showing the key results, and
3. a discussion of the implications of these results.

In the Conclusions section of our SLR, the following questions should be clearly reported:

- What, precisely, was the contribution of our SLR? (for both, researchers and practitioners)

- What was our new results?

- Why should the reader believe our results?

A generic, very frequent wording for the Conclusions section is the following:

*We have carried out a systematic literature review with the objective of <<Goal>>. An exhaustive search in online databases, journals and conferences, allowed the identification of a significant number of primary studies. <<XXX>> studies were selected, after applying the exclusion criteria. The following conclusions are based on data extracted from those XXX studies.*

It is advisable to follow an orderly sequence to report the conclusions for each RQ, and finally, the DQs (conclusions from Demographic Questions). For example:

- Conclusions from RQ1: <<key results>> <<possible explanation>>
- Conclusions from RQ2: <<key results>> <<possible explanation>>
- …
- Conclusions from RQn: <<key results>> <<possible explanation>>
- Conclusions from DQ1: <<key results>> <<possible explanation>>
- Conclusions from DQ2: <<key results>> <<possible explanation>>
- …
- Conclusions from DQm: <<key results>> <<possible explanation>>
- One or two general conclusions
- Possibly some proposals for future work

As future work, we intend to combine the evidence from this SLR with empirical data, obtained from a survey addressed to all industrial organizations identified in our set of selected papers. The questionnaire will focus on the current status of their SPLs, its evolution and the emergence of new benefits/disadvantages since the publication of the selected paper. Once the benefits and drawbacks are classified, we believe that they could be a useful tool for professionals to support their decision-making and a guide for academics to explore new research opportunities.

Finally, another important aspect in reporting is the choice of material to include as appendices. Some recommended appendices are:

- selected studies: <<an ordered list of all bibliographic references>>
- borderline studies: << the bibliographic references and the rationale for their exclusion>>
- rubrics for the SLR (template)
- self-assessment (applying the template)

## 3. Conclusions

This document contains the detailed protocol needed to perform a SLR. The authors have made an effort to include all the key points, which should be considered to ensure, as far as possible, the success of the study. We also include a list of key references that we believe will be useful for all those interested in the topic of EBSE and systematic secondary studies..

We will be grateful to readers willing to make comments that help improve this document.

### References

[1]     B.A. Kitchenham, D. Budgen, P. Brereton, Evidence-Based Software Engineering and Systematic Reviews, CRC Press, 2015.

[2]     K. Petersen, S. Vakkalanka, L. Kuzniarz, Guidelines for conducting systematic mapping studies in software engineering: An update, Inf. Softw. Technol. 64 (2015) 1–18. doi:10.1016/j.infsof.2015.03.007.

[3]     D. Budgen, P. Brereton, S. Drummond, N. Williams, Reporting systematic reviews: Some lessons from a tertiary study, Inf. Softw. Technol. 95 (2018) 62–74. doi:10.1016/j.infsof.2017.10.017.

[4]     M. Kuhrmann, M. Daniel, On the pragmatic design of literature studies in software engineering : an experience-based guideline, Empir. Softw. Eng. (2017) 2852–2891. doi:10.1007/s10664-016-9492-y.

[5]     F.J. der Linden, K. Schmid, E. Rommes, Software product lines in action: the best industrial practice in product line engineering, Springer Science & Business Media, 2007.

[6]     S.P. Gregg, R. Scharadin, P. Clements, The more you do, the more you save: The superlinear cost avoidance effect of systems product line engineering, ACM Int. Conf. Proceeding Ser. 20–24–July (2015) 303–310. doi:10.1145/2791060.2791065.

[7]     K. Schmid, M. Verlage, The economic impact of product line adoption and evolution, IEEE Softw. 19 (2002) 50–57. doi:10.1109/MS.2002.1020287.

[8]     J.M. Neighbors, Software construction using components, University of California, Irvine, 1980.

[9]     D. Badampudi, C. Wohlin, K. Petersen, Experiences from using snowballing and database searches in systematic literature studies, in: Proc. 19th Int. Conf. Eval. Assess. Softw. Eng., 2015: p. 17. doi:10.1145/2745802.2745818.

[10]    J. Bailey, C. Zhang, D. Budgen, M. Turner, S. Charters, Search Engine Overlaps : Do they agree or disagree?, Second Int. Work. Realis. Evidence-Based Softw. Eng. (REBSE '07). (2007) 2–2. doi:10.1109/REBSE.2007.4.

[11]    D. Budgen, P. Brereton, Performing systematic literature reviews in software engineering, Int. Conf. Soft. Engin. (2006) 1051. doi:10.1145/1134285.1134500.

[12]    L. Chen, M. Ali Babar, H. Zhang, Towards an evidence-based understanding of electronic data sources, in: Proc. 14th Int. Conf. Eval. Assess. Softw. Eng., 2010.

[13]    B. Kitchenham, Z. Li, A. Burn, Validating search processes in systematic literature reviews, in: Proceeding 1st Int. Work. Evidential Assess. Softw. Technol. EAST 2011, Conjunction with ENASE 2011, Beijing, 2011: pp. 3–9.

[14]    B. Kitchenham, P. Brereton, M. Turner, M. Niazi, S. Linkman, R. Pretorius, D. Budgen, The impact of limited search procedures for systematic literature reviews - A participant-observer case study, in: 3rd Int. Symp. Empir. Softw. Eng. Meas., 2009: pp. 336–345. doi:10.1109/ESEM.2009.5314238.

[15]    C. Wohlin, Guidelines for snowballing in systematic literature studies and a replication in software engineering, in: 18th Int. Conf. Eval. Assess. Softw. Eng. (EASE 2014), 2014: p. 38. doi:10.1145/2601248.2601268.

[16]  L. Northrop, Software Product Lines: Practices and Patterns, Addison-Wesley, 2002.

[17]  Product Line Hall of fame, Waseda Univ. Kishi Lab. (n.d.). http://splc.net/fame.html (accessed March 29, 2019).

[18]  M. Turner, Digital libraries and search engines for software engineering research: an overview (Technical report), 2010. https://community.dur.ac.uk/ebse/resources/notes/tools/SearchEngineIndex_v5.pdf.

[19]  B. Kitchenham, S. Charters, Guidelines for performing Systematic Literature Reviews in Software Engineering. Version 2.3, 2007. doi:10.1145/1134285.1134500.

[20]  K. Petersen, N. Bin Ali, Identifying Strategies for Study Selection in Systematic Reviews and Maps, 2011 Int. Symp. Empir. Softw. Eng. Meas. (2011) 351–354. doi:10.1109/ESEM.2011.46.

[21]  N. Bin Ali, K. Petersen, Evaluating strategies for study selection in systematic literature studies, in: Proc. 8th ACM/IEEE Int. Symp. Empir. Softw. Eng. Meas., 2014: p. 45. doi:10.1145/2652524.2652557.

[22]  L. Chen, M. Ali Babar, A systematic review of evaluation of variability management approaches in software product lines, Inf. Softw. Technol. 53 (2011) 344–362. doi:10.1016/j.infsof.2010.12.006.

[23]  I. Steinmacher, A.P. Chaves, M.A. Gerosa, Awareness support in distributed software development: A systematic review and mapping of the literature, Comput. Support. Coop. Work. 22 (2013) 113–158.

[24]  A. Ampatzoglou, S. Bibi, P. Avgeriou, M. Ver-beek, A. Chatzigeorgiou, Identifying, categorizing and mitigating threats to validity in software engineering secondary studies, Inf. Softw. Technol. (2018).

[25]  K. Petersen, C. Gencel, Worldviews, research methods, and their relationship to validity in empirical software engineering research, in: Softw. Meas. 2013 Eighth Int. Conf. Softw. Process Prod. Meas. (IWSM-MENSURA), 2013 Jt. Conf. 23rd Int. Work., 2013: pp. 81–89. doi:10.1109/IWSM-Mensura.2013.22.

[26]  H. Zhang, M.A. Babar, P. Tell, Identifying relevant studies in software engineering, Inf. Softw. Technol. 53 (2011) 625–637.